# First Principles Study of SnO Under High Pressure

M. A. ALI[†,ϒ,*], A. K. M. A. ISLAM [ϒ,‡], N. JAHAN[†], S. KARIMUNNESA[†]

[†]*Department of Physics, Chittagong University of Engineering and Technology, Chittagong-4349, Bangladesh.*

[ϒ]*Department of Physics, University of Rajshahi, Rajshahi-6205, Bangladesh.*

[‡]*International Islamic University Chittagong, 154/A College Road, Chittagong, Bangladesh.*

This article reports the first-principles study of SnO under high pressure within the generalized gradient approximation (GGA). We have calculated the structural, elastic, electronic and optical properties of SnO. The elastic properties such as the elastic constants $C_{ij}$, bulk modulus, shear modulus, Young modulus, anisotropic factor, Pugh ratio, Poisson's ratio are calculated and analyzed. Mechanical stability of SnO at all pressure is confirmed using Born stability conditions in terms of $C_{ij}$. It is also found that SnO exhibits very high anisotropy. The energy band structure and density of states are also calculated and analyzed. The results show the semiconducting and metallic properties at 0 (zero) and high pressure, respectively. Furthermore, the optical properties are also calculated. All the results are compared with those of the SnO where available but most of the results at high pressure are not compared due to unavailability of the results.

## 1. Introduction

In recent times, research based on high-pressure has undergone a new start, permits substantial developments of high-pressure research with available experimental facility by means of advances in techniques and instrumentation.[1] High-pressure research has drawn huge concentration from researchers due to their vital use to discover the new types of physical behavior and new potential candidate materials for technological application. It has also impact on the exploring of previously unexplored chemical and physical properties. Researchers from many fields of natural sciences like Physics, molecular Chemistry, Geophysics, and Biology, have found an enormous impact on their respective fields.[2-4] To explore and develop the properties of materials under high pressure it is necessary to identify the different strategies.

SnO is found to be crystallized with tetragonal litharge structure (*P4/nmm, S.G. 129, Z* = 2). The oxygen atoms and Sn atoms placed at the Wyckoff position of 2a and 2c site, respectively.[5] The tin oxides (SnO and $SnO_2$) are suitable candidate materials for thin-films processing and applications. Films of these oxides have become leading important due to their potential applications in solar cells,[6] heat reflectors for advanced glazing in solar applications[7,8] and as various gas sensors.[9–12]

Due to the variety of applications mentioned above, many researchers have paid attentions which are reported in recent years.[13–24] Specially, $SnO_2$, have been largely studied because of its n-type semiconducting nature with wide band gap (3.6 eV), while research work on SnO are seldom reported due to its decomposing tendency at high temperature.[25,26] But during the last few years researchers have paid their attention. Some results of experimental and theoretical studies on SnO have been reported. High pressure study of electronic and structural properties of SnO up to 10 GPa has been studied by Christensen *et al.*[20] Giefers *et al*[21] has investigated the structural properties of SnO at high pressure up to 51 GPa. The pressure-induced phase transition in SnO has been studied by Lie et al.[22] L. A. Errico[23] has studied the structural and electronic properties of SnO and $SnO_2$. The $SnO_2$ and SnO were investigated by Q J Liu *et al*.[24] First principles studies of SnO at different structures have also been investigated by Erdem *et al*.[25]

It is reported, at zero pressure SnO is a semiconductor but transforms to a metal at 5 GPa.[26] Christensen *et al*[20] reported that a change in the slope of *c/a* curve as a function pressure was found around 5 GPa which is responsible for that transition. Recently, a superconducting state of SnO at a pressure, p > 6 GPa is reported[27] where a maximum $T_c$ was found 1.4 K at 9.3 GPa. Though a phase transition of *α*-SnO to *γ*–SnO under pressure is reported[28–30] but such type of transition is not observed by the recent experiments carried out by Wang *et al*[26] and Giefers *et al.*[21] It was concluded about the phase transition that this might be induced by non- hydrostatic conditions. This motivates us to perform a complete study of SnO under hydrostatic pressure. However, as per of our knowledge complete study of SnO under high pressure have not been done yet.

---

[†]M. A. ALI. Department of Physics, Chittagong University of Engineering and Technology, Chittagong-4349, Bangladesh.
[‡]A. K. M. Azharul Islam. International Islamic University Chittagong, 154/A College Road, Chittagong, Bangladesh.
[*]Corresponding author: M. A. ALI, E-mail: ashrafphy31@gmail.com

Therefore, in this article, a systematic investigation of structural, elastic, electronic and optical properties of SnO under high pressure is presented. We expect that our study would add novel information on the existing properties.

## 2. Computational Methods

The CAmbridge Serial Total Energy Package (CASTEP) code[31] is used to perform the calculation of properties of SnO. The method used in this code is the plane wave pseudopotential approach based on the density functional theory (DFT).[32] During the calculations the exchange-correlation potential are treated within the GGA using Perdew-Burke-Ernzerhof (PBE).[33] A 6 × 6 × 5 k-point mesh of Monkhorst-Pack scheme[34] was used for integration over the first Brillouin zone. The convergence of the plane wave is done with a kinetic energy cutoff of 500 eV. Excellent convergence is guaranteed by testing the Brillouin zone sampling and the kinetic energy cutoff which make the tolerance for self-consistent field, energy, maximum force, maximum displacement, and maximum stress to be $5.0\times10^{-7}$ eV/atom, $5.0\times10^{-6}$ eV/atom, 0.01 eV/Å, $5.0\times10^{-4}$ Å, and 0.02 GPa, respectively. The optimized structure of SnO was obtained using the Broydon-Fletcher-Goldfarb-Shanno (BFGS) minimization technique.[35]

## 3. Results and discussion

### 3.1. *Structural properties and their pressure dependence*

Figure 1 shows the crystal structure of tin monoxide (SnO), the tetragonal litherge-type structure (SG P4/nmm, SG No. 129); in which the position of oxygen atoms are (0,0,0) and (0,0.5,0) of Wyckoff 2a site while the position of Sn atoms are (0.5,0,u) and (0,0.5,1−u) of 2c site; where the internal parameter u = 0.2361, is a measure of the position of the O atom along the c-axis. The total energy is minimized by the geometry optimization of the SnO at zero pressure using CASTEP code. After geometry optimization, the lattice parameters of SnO are taken at different pressures. The optimized lattice parameters are given in Table 1. Our calculated results of SnO are compared with the available results.[21,24] The calculated lattice parameters are in reasonable agreement with the experimental results. Considering the nature of GGA-PBE functional, it is not surprising that theoretical estimates of the lattice parameters and cell volume are slightly higher than the experimental ones (Table 1).

Table 1. Lattice constants, *a*, *c*, *c/a* and volume, *V* of SnO.

| Compound | Crystal system | a | c | c/a | V | |
|---|---|---|---|---|---|---|
| | | 3.90 | 4.95 | 1.27 | 75.95 | This Cal. |
| SnO | Tetragonal | 3.80 | 4.83 | 1.27 | 69.97 | Expt.[21] |
| | | 3.84 | 4.91 | 1.27 | 72.72 | Theo.[24] |

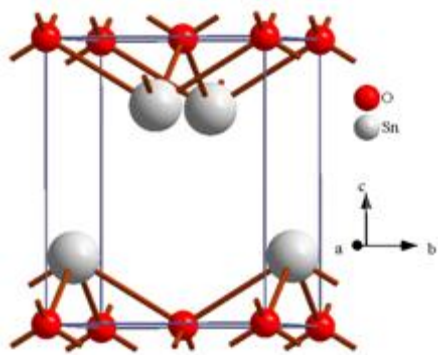

Fig. 1. The SnO litharge-type tetragonal structure.

Figure 2 shows the normalized values $a/a_0$, $c/c_0$, $c/a$ and $V/V_0$ of SnO (where $a_0$, $c_0$ and $V_0$ are the structural parameters at zero pressure) for different pressure. It is seen that the volume decreases with increasing pressure. Since the tin monoxide is soft materials, therefore it is expected to shrink the volume of the cell when it is pressurized uniformly. Moreover, the normalized lattice constant $c/c_0$ decreasing more quickly than $a/a_0$ with increasing pressures indicating the larger compression along c-axis. A change in the slope of c/a ratio is found around 5 GPa [Fig. 2 (b)], which is reported as an indication of semiconductor to metal transition.[20] Our calculated results are matched well with the results obtained by Giefers *et al.*[21] The interlayer compressibility, $d(\ln c)/dP$ = 0.01773312 GPa$^{-1}$ is about 4.92 times higher than the basal plane linear compressibility $d(\ln a)/dP$ = 0.00362235 GPa$^{-1}$, also indicates the larger response to the pressure of the lattice constant *c* than *a*.

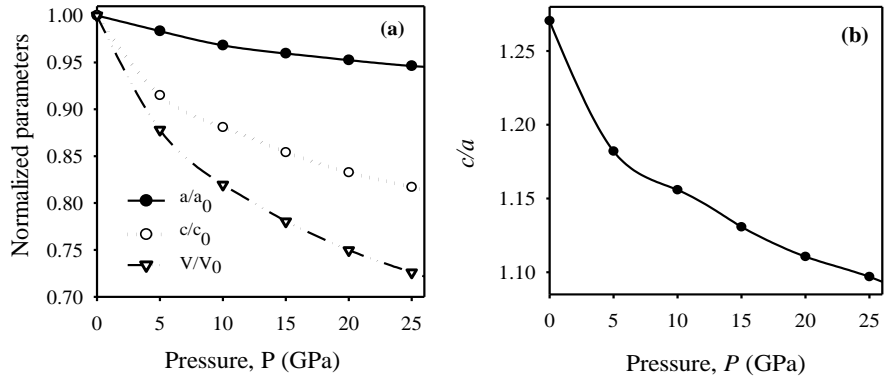

Fig. 2. The pressure dependence of the normalized lattice parameters, (a) $a/a_0$, $c/c_0$ and $V/V_0$ (b) $c/a$ of SnO.

### 3.2. *Elastic properties and their pressure dependence*

The independent elastic constants, $C_{ij}$ of SnO are calculated in the pressure range 0 to 25 GPa and are given in Table 2 along with other results for comparison.[25] As per of our knowledge, the elastic constants with limited pressure (0 to 1.5 GPa) have been reported[22], but for such long pressure range (0 to 25 GPa) is not calculated before this study. Therefore the comparisons of results obtained at high pressure are not possible at this time. The finite strain technique has been used to study the elastic properties of SnO. In this case a homogeneous deformation (strain) is applied and resulting stress is calculated. A compound is considered to be mechanically stable if the values of $C_{ij}$ are all positive and also satisfy the conditions for mechanical stability of tetragonal crystals:[36] $C_{11} > 0$, $C_{33} > 0$, $C_{44} > 0$, $C_{66} > 0$, $(C_{11} - C_{12}) > 0$, $(C_{11} + C_{33} - 2C_{13}) > 0$ and $[2(C_{11} + C_{12}) + C_{33} + 4C_{13}] > 0$. It is found that (Table 2) these conditions are satisfied by SnO and hence mechanically stable.

Table 2. The elastic constants $C_{ij}$ (GPa), bulk modulus, *B* (GPa), shear modulus, *G* (GPa), Young's modulus, *Y* (GPa), Poisson ratio, *v*, and *G/B* of SnO under different pressure.

| P (GPa) | $C_{11}$ (GPa) | $C_{12}$ (GPa) | $C_{13}$ (GPa) | $C_{33}$ (GPa) | $C_{44}$ (GPa) | $C_{66}$ (GPa) | B (GPa) | G (GPa) | Y (GPa) | v | G/B | Ref. |
|---|---|---|---|---|---|---|---|---|---|---|---|---|
| 0 | 92 | 83 | 24 | 35 | 21 | 72 | 43 | 20 | 52 | 0.29 | 0.46 | [This] |
|  | 147.8 | 83.9 | 53.4 | 102.2 | 43.9 | 74.6 | 86.8 |  |  |  |  | [25] |
|  | 50 | 38 |  |  | 20 | 80 | 31 |  |  |  |  | [22] |
|  |  |  |  |  |  |  | 50 |  |  |  |  | [28] |
| 5 | 131 | 104 | 39 | 76 | 41 | 87 | 71 | 37 | 94 | 0.27 | 0.52 |  |
| 10 | 154 | 126 | 58 | 115 | 57 | 96 | 96 | 45 | 116 | 0.29 | 0.46 |  |
| 15 | 186 | 155 | 91 | 147 | 73 | 105 | 128 | 51 | 135 | 0.32 | 0.39 |  |
| 20 | 204 | 173 | 107 | 167 | 80 | 112 | 145 | 54 | 144 | 0.33 | 0.37 |  |
| 25 | 231 | 188 | 119 | 178 | 96 | 119 | 160 | 63 | 167 | 0.32 | 0.39 |  |

The bulk modulus *B*, shear modulus *G*, Young's modulus *Y*, the Poisson ratio *v* and Pugh ratio are also calculated and given in Table 2. *Y* and *v* are calculated from *B* & *G* by using the relationships: $Y = 9BG/(3B + G)$, $v = (3B-Y)/6B$,[37,38] where B and G are obtained from the Voigt-Reuss-Hill (VRH) average scheme.[39,40]

The elastic constants of SnO under pressure (up to 25 GPa) are summarized in Table 2. The elasticity in length is represented by elastic constant $C_{11}$ ($C_{33}$) which is the consequence of longitudinal strain, whereas the shearing elasticity is related to the elastic constants $C_{12}$ ($C_{13}$) and $C_{44}$ ($C_{66}$). Figure 3(a) shows the pressure dependence of $C_{11}$, $C_{33}$, $C_{66}$, $C_{12}$, $C_{13}$ and $C_{44}$. The variation of elastic constants $C_{44}$, $C_{66}$ show lesser sensitivity to pressure compared to other elastic constants.

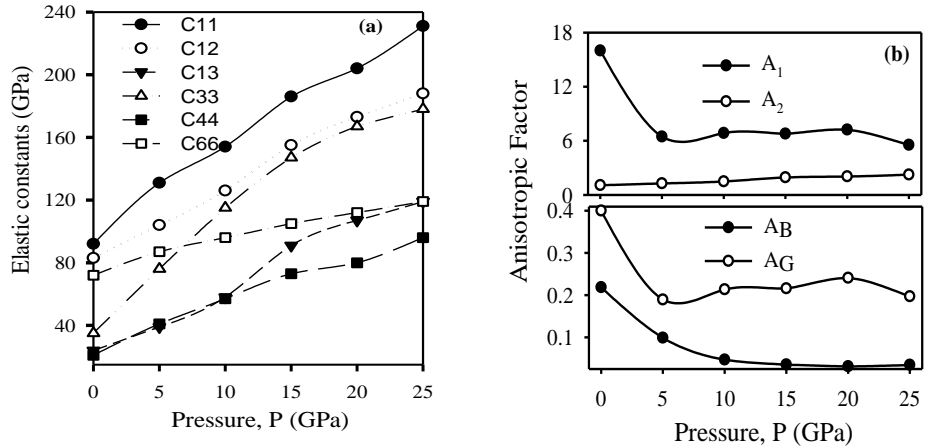

Fig. 3. (a) The pressure dependence of elastic constants $C_{ij}$ (GPa) and (b) anisotropic factor of litharge SnO as function of pressure.

Now let us consider the other elastic constants presented in Table 2. The bulk modulus describes the material's response to uniform pressure i.e., the resistance to change in volume. We see that it is more difficult to change the volume of SnO at higher pressure because the values ($B$) are found to be increased with pressure. The shear modulus $G$ represents the resistance to elastic shear deformation and is also found to be increased at high pressure phases. The stiffness of materials is generally expressed by its Young's modulus $Y$. The higher value of $Y$ indicates higher stiffness of the material. It is also found that the phase under consideration is found to be stiffer at higher pressure (see table 2). The values of $C_{11}$, $C_{33}$, $C_{66}$, $C_{12}$, $C_{13}$, $C_{44}$, B, G, and Y do not indicate the measurement of the mechanical hardness of the materials but these are normally greater for comparatively harder materials. It is found that the constants mentioned above are found to be increased with the increase of pressure. Therefore, we can conclude that the SnO becomes harder with rise in pressure. Brittle/ductile nature of solid can be predicted in terms of Poisson's ratio and Pugh ration. The value of Poisson's ratio, $v$ is typically 0.33 for ductile metal and for brittle material its value is small[41]. It is found that the SnO shows ductile metallic behavior at high pressure but lies in borderline of brittle to ductile at low pressure. Pugh[42] (Pugh ratio, G/B) gave a critical value for ductile-brittle transition. For brittle materials $G/B$ ratio is greater than 0.5 while for ductile materials its value is less than 0.5. From Table 2, we see that SnO becomes more ductile with increasing pressure while at low pressure (even zero pressure) it lies in borderline of brittle to ductile transition.

Table 3. Different anisotropic constants of SnO as a function of pressure.

| Pressure, P (GPa) | A | $B_v$ (GPa) | $B_R$ (GPa) | $G_V$ (GPa) | $G_R$ (GPa) | $A_B$ | $A_G$ | $A_1$ | $A_2$ |
|---|---|---|---|---|---|---|---|---|---|
| 0 | 4.666667 | 53 | 34 | 28 | 12 | 0.218 | 0.4 | 16 | 1.06 |
| 5 | 3.037037 | 78 | 64 | 44 | 30 | 0.098 | 0.189 | 6.4 | 1.27 |
| 10 | 4.071429 | 101 | 92 | 54 | 35 | 0.046 | 0.213 | 6.8 | 1.49 |
| 15 | 4.709677 | 133 | 124 | 62 | 40 | 0.035 | 0.215 | 6.7 | 1.93 |
| 20 | 5.16129 | 150 | 141 | 67 | 41 | 0.030 | 0.240 | 7.2 | 2.03 |
| 25 | 4.465116 | 166 | 155 | 76 | 51 | 0.034 | 0.196 | 5.5 | 2.24 |

Anisotropic factor of any crystal is very significant in applied science and also in crystal physics. Therefore, it is very important to describe properly the anisotropic behavior of any materials for engineering application. The anisotropy of atomic bonding in different planes is described by the shear anisotropic factors which are defined by following equations for tetragonal litharge type SnO for the {1 0 0} and {0 0 1} shear planes:[43]

$$A_1 = \frac{2C_{66}}{C_{11} - C_{12}}; \quad A_2 = \frac{4C_{44}}{C_{11} + C_{33} - 2C_{13}}.$$

In general for an isotropic crystal, the factors $A_1$ and $A_2$ must equate to one, while departure from unity denotes the degree of elastic anisotropy. The anisotropic factors $A_1$ and $A_2$ as a function of pressure are plotted in Fig. 3. It is found that $A_2$ almost keeps constant and $A_1$ decreases sharply up to pressure 5 GPa and becomes almost constant up to 20 GPa and thereafter decreases with pressure. The anisotropy only depends on the symmetry of the crystal. A change in crystal structure under applied pressures has been occurred due to the variations of lattice constants *a* and *c* at various pressures. As a result, the elastic anisotropy is also changed because of the pressure dependence of elastic constants. The elastic anisotropy can be defined in terms of bulk and shear modulus as, $A_B = (B_V - B_R)/(B_V + B_R)$ and $A_G = (G_V - G_R)/(G_V + G_R)$. The materials show isotropic nature when the values of $A_B$ ($B_V = B_R$) and $A_G$ ($G_V = G_R$) become zero, while a non-zero value indicates the anisotropy of the materials. The calculated $A_B$ and $A_G$ are also presented as a function of pressure in Fig. 3. As the pressure increases, the anisotropy of the bulk modulus $A_B$ decreases exponentially whereas the shear modulus anisotropy $A_G$, decreases dramatically with pressure.

### 3.3. *Electronic properties at 0 GPa and 5 GPa*

The band structures of SnO along high-symmetry directions of the crystal Brillouin zone (BZ) are displayed in Fig. 4 (a and b) for 0 GPa and 5 GPa only. The Fermi level is set at 0 (zero) eV in the energy scale. No significant changes are observed in the band structures at higher pressure which are not shown in figures. The band structure shows that the SnO is an indirect band gap semiconductor. The band gap of 0.4598 eV is found from Γ to M [Fig. 4(a)]. The minimum direct band gap is found to be 2.0 eV. Our results can be compared with the values of 0.3 eV (indirect) and 2.0 eV (direct) by Errico[23] using the Full-Potential Linearized Augmented-Plane Wave (FP-LAPW) method as implement in WIEN97 code and also with the values obtained by Liu et al[24] 0.476 eV (indirect) and 2.0 eV (direct), by using LDA as implemented in CASTEP code. SnO is normally known as a semimetal or a small gap semiconductor with an optical gap of 2.5-3 eV,[44] where as the fundamental gap of 0.7 eV[45] is also reported. It is important to note that the pressure has enormous effect on the band gap of SnO and the band gap very much responsive to the c/a ratio.[20,23] Normally, the values of band gap energy obtained from DFT calculation are smaller than the experimental values. We have pointed out this fact briefly in another article[18] which is based on $SnO_2$.

The total and partial DOS for 0 & 5 GPa are shown in Fig. 5. The details study of electronic structure can be obtained from the total and partial (DOS) of SnO. The contribution from O-*p* states is attributed to the band structure around the Fermi level, $E_F$. In order to describe the contribution from different states we sub-divide the band below Fermi level into three energy region. The lower band around -18 eV to -20 eV is mainly derived from O-*s* states while the little contribution from S-*s* and S-*p* states also noticeable. The valence band from 0 to around -9 eV can be divided into two sub-bands and namely upper sub-band and lower sub-band. The upper sub-band is derived mainly from O-*p* states down to around – 7.5 eV when a small contribution of Sn-*p* states is noticeable and the lower sub-band is derived from Sn-*s* states. The DOS of the conduction bands are mainly derived from Sn-*p* states while Sn-*s* and O-*p* states also have a little contribution.

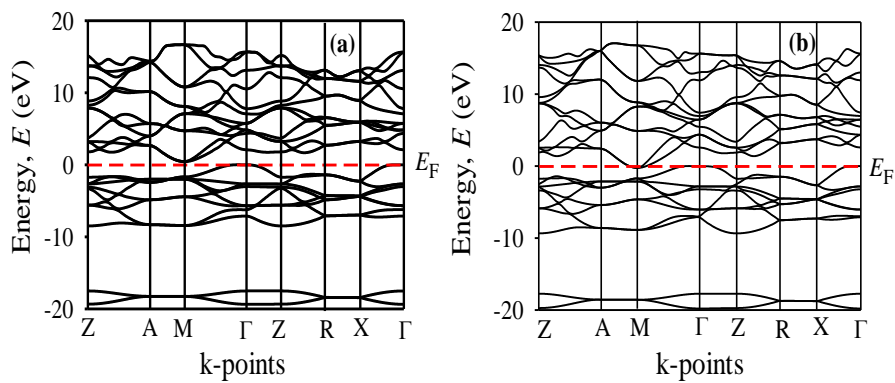

Fig. 4. Band structure at (a) 0 GPa and (b) 5 GPa.

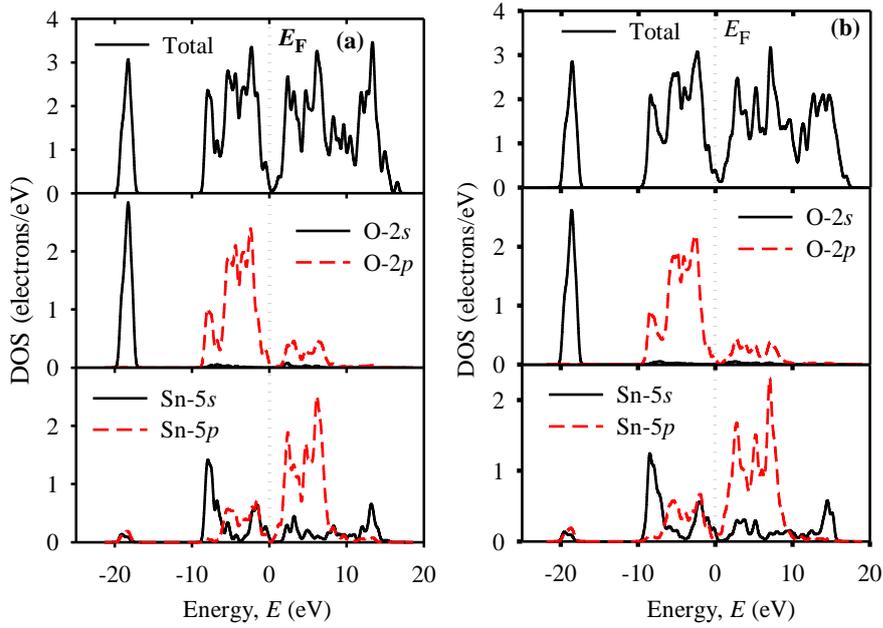

Fig. 5. Density of states (DOS) at (a) 0 GPa and (b) 5 GPa.

To study the pressure effects on electronic properties we now compare the band structures and density of DOS of SnO for 0 and 5 GPa as displayed in Figure 4 and 5. Figure 4 & 5 clearly distinguish the bands and DOS by the existence of additional features obtained at 0 & 5 GPa due to the effect of hydrostatic pressure. No bands cross the Fermi level in the band-structure [Figure 4(a)] obtained at 0 GPa but the appearance of extra bands in band gap region is the main feature due the pressure effects. From Fig. 4(b), it is clearly seen that the conduction band cross significantly the Fermi level which is the evidence for the overlapping of valence band and conduction band. Though the valence band and conduction band does not overlap directly because of the indirect band gap of SnO. This can be explained in terms of partial density of states. In Fig. 5 (b) we can see an additional peak of Sn-*s* states at the Fermi level. Also the contribution from Sn-*p* states at Fermi level is increases with increasing pressure. Due to hybridization of Sn-*p* and Sn-*s* states at Fermi level at higher pressure, SnO transform into metal from semiconductor. If we further increase the pressure no significant change have been observed but only the value of DOS at Fermi is found to be slightly increased.

### 3.4. *Optical properties at* 0 *GPa and 5 GPa*

When an electromagnetic radiation is incident on the materials, different materials behave in different way. The complex dielectric functions; $\varepsilon(\omega) = \varepsilon_1(\omega) + i\varepsilon_2(\omega)$ is one of the main optical characteristics of solids and the other optical constants can be obtained from it. The imaginary part $\varepsilon_2$ ($\omega$) is calculated in CASTEP[46] numerically by a direct evaluation of the matrix elements between the occupied and unoccupied electronic states. The expression for the $\varepsilon_2$ ($\omega$) can be found elsewhere.[47,48] The Kramers-Kronig relations are used to derive the real part $\varepsilon_1(\omega)$ of dielectric function. The other optical constants described in this section are derived from $\varepsilon_1(\omega)$ and $\varepsilon_2(\omega)$ using the equations given in ref. 46.

The optical constants of SnO are shown in Fig. 6 (left and right panel) for (100) polarization direction. To smears out the Fermi level for effective k-points on the Fermi surface, we used a 0.5 eV Gaussian smearing. Similar to the electronic properties only 0 and 5 GPa data are shown because of the insignificant change in optical properties after 5 GPa.

The electronic properties of crystalline material are mainly characterized by the imaginary part, $\varepsilon_2(\omega)$ of dielectric function, $\varepsilon(\omega)$ which depicts the probability of photon absorption. The peaks of $\varepsilon_2(\omega)$ is associated with the electron excitation. There are three peaks at 3.5, 7.8, and 15.7 eV [Fig. 6 (left panel, b)]. The main peak at 4.5 eV may arise due to the electrons transition from O-2*p* and Sn-5*p* states. The second peak at 8 eV caused by the electrons transition from Sn-5*s* states. The positions of the peaks are in good agreement with Liu *et al*.[24] At higher pressure (5 GPa) the peaks positions is less like to be remain same. The dielectric constant $\varepsilon_1$ (0) [Fig.

6 (left panel, a)] at zero energy has the value of about 7, indicating the dielectric nature of SnO. The position of the peak (4.5 eV) is also in good agreement with Liu et al.[24] At high pressure the large negative value of $\varepsilon_1$ is a indication of Drude-like behavior which is the characteristics of metals.

The refractive index is another technically important parameter for optical materials. The spectra for refractive index $n$ and the extinction coefficient $k$ are demonstrated in Fig. 6 (left panel, c & d). The static value of $n(0)$ is obtained of about ~ 2.8 at 0 GPa which is comparable with that obtained by Liu et al.[24] while the value is found to be 3.3 at 5 GPa. The peaks positions and pattern of curves for refractive index and extinction coefficient are similar to the curve obtained by Liu et al.[24]

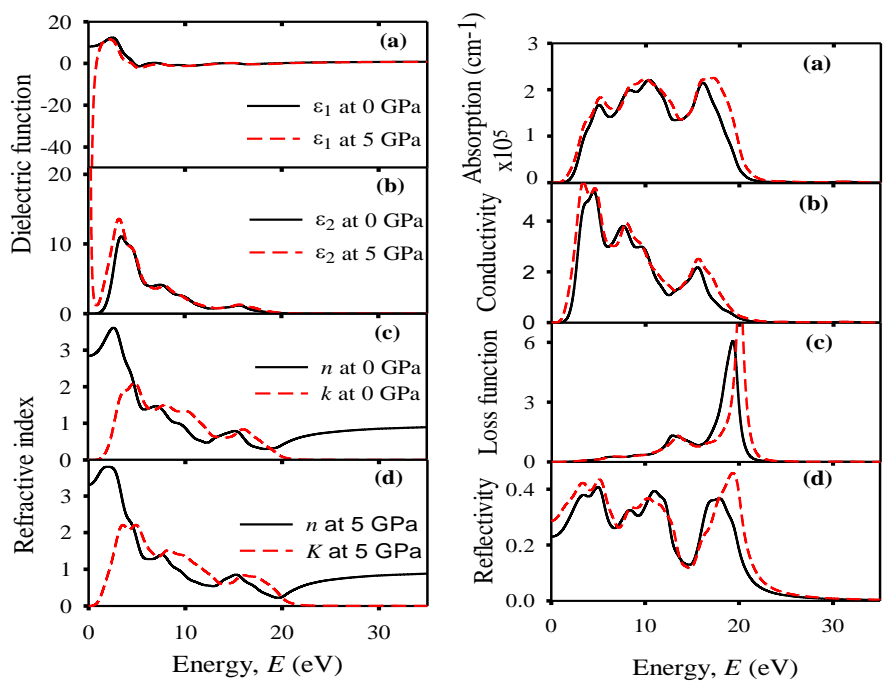

Fig. 6. Left panel: (a) $\varepsilon_1$, (b) $\varepsilon_2$, (c) n & k at 0 GPa, (d) n & k at 5 GPa; Right panel: (a) absorption coefficient, (b) photoconductivity, (c) loss function, (d) reflectivity of SnO.

Figure 6 (right panel, a & b) shows the absorption coefficient and conductivity spectra of the compound under consideration at 0 and 5 GPa. The curves for 0 GPa delayed to start due the existence of band gap but started in well advance for 5 GPa compared to 0 GPa showing the reduction of band gap. The results are matched well with bands structure results. There is no significant change in the positions and height of the peaks due to applied pressure after 5 GPa. The curve is associated with several peaks within the energy range studied.

The loss function $L(\omega)$, shown in Fig. 6 (right panel, c), is defined as energy loss of a electron with high velocity passing through the materials. This curve is characterized by a peak which is known as bulk plasma frequency $\omega_P$, which occurs at $\varepsilon_2 < 1$ and $\varepsilon_1 = 0$. In Fig. 6 (right panel, c), The value of the effective plasma frequency $\omega_P$ is found to be ~ 19.3 and 20 eV for 0 and 5 GPa, respectively. The material becomes transparent, when the frequency of incident photon is greater than $\omega_P$. The reflectivity curve is shown in Fig. 6 (right panel, d). It is found that the reflectivity of the compound starts with a value of ~ 0.23 & 0.28 for 0 and 5 GPa, respectively. After an increase with photon energy up to ~ 5 eV, the reflectivity falls again. The value of reflectivity finally rises to reach a maximum value of ~ 0.35 and 0.48 at ~ 18 and 20 eV for 0 and 5 GPa, respectively. The high value of reflectivity at 5 GPa compared to the value obtained at 0 GPa in very low energy range is a sign of the increase in conductivity at higher pressure. Moreover, the peak of loss function is associated with the trailing edges of the reflection spectra.

## 4. Conclusion

We have studied the structural, elastic, electronic and optical properties of SnO under high pressure. The lattice constant $c$ is found to be more responsive to pressure than $a$. The change in the $c/a$ ratio might be attributed to the semiconductor to metal transition. The SnO is found to be mechanically stable. The elastic constant are found to be increased with increasing pressure results an improvement of mechanical hardness. Also the SnO becomes more ductile at higher pressure. The band structures at 0 (zero) GPa shows the semiconducting behavior but at a higher pressure a transition to metal is found. Finally, optical properties have also been calculated and discussed. The optical properties also distinct the transition from semiconductor to metal at zero pressure and high pressure, respectively.